\documentclass[aps,pra,twocolumn,a4paper,
 superscriptaddress,
 floatfix,10pt,longbibliography]{revtex4-1}
\usepackage[utf8]{inputenc}
\usepackage{amsmath}
\usepackage{amsfonts}
\usepackage{amssymb}
\usepackage{graphicx}
\usepackage[T1]{fontenc}
\usepackage{float}
\usepackage{booktabs}
\usepackage{comment}
\usepackage[dvipsnames]{xcolor}
\usepackage{braket}
\usepackage{bm}
\usepackage{hyperref}

\graphicspath{{Figures/}{./}}

\newcommand{\Sent}{\ensuremath{S_\mathrm{ent}}}
\definecolor{OxfBlue}{rgb}{0, 0.333, 0.710}

\begin{document}

\title{Characterizing the phase diagram of finite-size dipolar Bose-Hubbard systems}

\author{Paolo Rosson}
\affiliation{Clarendon Laboratory, University of Oxford, Parks Road, Oxford OX1 3PU, United Kingdom}

\author{Martin Kiffner}
\affiliation{Centre for Quantum Technologies, National University of Singapore, 3 Science Drive 2, Singapore 117543}
\affiliation{Clarendon Laboratory, University of Oxford, Parks Road, Oxford OX1 3PU, United Kingdom}

\author{Jordi Mur-Petit}
\email{Corresponding author: jordi.murpetit@physics.ox.ac.uk}
\affiliation{Clarendon Laboratory, University of Oxford, Parks Road, Oxford OX1 3PU, United Kingdom}

\author{Dieter Jaksch}
\affiliation{Clarendon Laboratory, University of Oxford, Parks Road, Oxford OX1 3PU, United Kingdom}
\affiliation{Centre for Quantum Technologies, National University of Singapore, 3 Science Drive 2, Singapore 117543}

\date{\today}

\begin{abstract}
We use state-of-the-art density matrix renormalization group calculations in the canonical ensemble to determine the phase diagram of the dipolar Bose-Hubbard model on a finite cylinder. 
We consider several observables that are accessible in typical optical lattice setups  and assess how well these quantities perform as order parameters.
We find that, especially for small systems, the occupation imbalance is less susceptible to boundary effects than the structure factor in uncovering the presence of a periodic density modulation.
By analysing the non-local correlations, we find that the appearance of supersolid order is very sensitive to boundary effects, which may render it difficult to observe in quantum gas lattice experiments with a few tens of particles.
Finally, we show that density measurements readily obtainable on a quantum gas microscope allow distinguishing between superfluid and solid phases using unsupervised machine-learning techniques.
\end{abstract}

\maketitle
\section{Introduction} \label{Intro}

The milestone observation of the superfluid-Mott insulator transition of the Bose-Hubbard model with ultracold atoms in an optical lattice~\cite{jaksch1998cold,Greiner2002} sparked a revolution in our approach to studying strongly-correlated quantum systems.
Experimental advances since then in the control and measurement of cold atomic gases have led to remarkable observations, among which are 
the development of quantum gas microscopes for both bosons and fermions~\cite{Bakr2009, Cheuk2015, sherson2010single}.

As a result, cold atomic systems are generally regarded as the go-to setting to study the competition between kinetic energy and interaction in strongly correlated systems because of their high degree of tunability and precision~\cite{lewenstein2012ultracold}.
However, interactions between neutral alkali atoms are usually of very short range. This precludes their use in studying systems with long-range interactions, for which theoretical calculations point to the existence of exotic phases including supersolid order~\cite{Capogrosso2010, zhang2018equilibrium, yamamoto2012quantum}, and spin-glass~\cite{sanpera2004atomic} and spin-ice \cite{Glaetzle2014} phases.
A number of experimental platforms are being developed to study long-range strongly-correlated phases, from atomic ions in Penning traps~\cite{britton2012engineered, schneider2012experimental}, to systems with dipole-dipole interactions (DDI) such as Rydberg atoms~\cite{saffman2010quantum, low2012experimental, gunter2013observing}, highly-magnetic atoms~\cite{de2013nonequilibrium, lu2011strongly, Bottcher2019,Tanzi2019,Chomaz2019}, or ultracold molecules~\cite{yan2013observation, hazzard2014many, Blackmore2019}.

An important question arising in this context is how these exotic phases can be experimentally detected in finite, frequently small ($N\sim 10^2$ particles), systems.
In theoretical calculations, these phases are generally identified, in the grand-canonical ensemble, through order parameters -- observables whose values undergo a sharp change when crossing a phase transition in the thermodynamic limit ($N\to\infty$).
Typical examples readily available in computations are the static structure factor to detect periodic density modulations characteristic of solids or the superfluid density or superfluid stiffness to detect superfluidity.
However, it is unclear how these phases appear in finite samples with a fixed number of particles (canonical ensemble), and what the best observables are to uncover them with the toolbox of quantum gas experiments.

Here, using density matrix renormalization group (DMRG) calculations on two-dimensional (2D) finite systems with sizes accessible to current experiments,
we present a systematic analysis of several experimentally available observables as candidate order parameters
to capture the phases of a paradigmatic model with long-range interactions, the dipolar Bose-Hubbard model~\cite{Capogrosso2010}.
We consider the occupation imbalance~\cite{schreiber2015observation,landig2016quantum, Tai2017}, and show that it can reliably predict the appearance of a periodic density modulation (a `density wave', DW).
In addition,
superfluidity (SF) is ascertained by establishing the algebraic decay of correlations in the one-body density matrix~\cite{Holzmann2007, Holzmann2007pnas, PitaBook}.
More generally, the entanglement entropy appears as the most flexible tool to detect phase transitions~\cite{Osborne2002, Vidal2003, Eisert2010, islam2015measuring}.
This set of observables allows us to identify a supersolid phase in the system~\cite{Capogrosso2010, zhang2018equilibrium, yamamoto2012quantum}, which we find to be very sensitive to finite-size effects.

Finally, motivated by recent work to identify phases with machine-learning techniques~\cite{wang2016discovering,hu2017discovering,dunjko2018machine}, we compare our microscopic calculations with an approach based on unsupervised learning. We find that a principal component analysis (PCA) of (simulated) experimental density measurements with single-site resolution is able to pinpoint the transition into density-ordered phases, but cannot distinguish a DW-ordered system from a supersolid, as it has no access to phase-coherence information.

This paper is organised as follows.
In Sec.~\ref{sec:model}, we introduce the dipolar Bose-Hubbard Hamiltonian, and briefly summarise our numerical approach to determine its ground state.
In Sec.~\ref{sec:ord-param} we present our numerical results benchmarking the candidate order parameters to detect density modulations (Sec.~\ref{sec:dw}), off-diagonal long-range order (Sec.~\ref{sec:sf}) and the entanglement entropy (Sec.~\ref{sec:ent-entropy}) in a small lattice system at half filling.
After cross-validating the various order parameters, in Sec.~\ref{sec:doped}  we consider doping the system with a single particle above half-filling and look for  evidence of supersolid order.
Finally, we compare the approach to phase identification based on order parameters with a machine-learning strategy utilising single-site-resolved density measurements in Sec.~\ref{sec:ml}.
We conclude with a discussion of our findings in Sec.~\ref{sec:conclusions}.

\section{Model and methods\label{sec:model}}

The dipolar Bose-Hubbard (dBH) model is described by the Hamiltonian
\begin{equation}\label{eq:Ham-with-U}
 H =-J\sum_{<i,j>} b_i^\dagger b_j 
 + U \sum_{i} n_i (n_i-1)
 + \dfrac{1}{2}\sum_{i\neq j} \dfrac{V}{r_{ij}^3}n_i n_j \, ,
\end{equation}
where $b_i$ is the annihilation operator for a spinless boson at a site labelled by the index $i$, satisfying bosonic commutation relations $[b_i, b_j^\dagger]=\delta_{ij}$,
$J$ is the hopping amplitude between nearest-neighbour sites,
$U$ is the on-site interaction energy,
$n_{i}=b^{\dagger}_{i}b_{i}$ is the number operator at site $i$.
$V$ characterizes the long-range static DDI, where the dipole moments of the molecules are aligned and orthogonal to the lattice~\cite{yan2013observation, hazzard2014many, Blackmore2019};
it can be expressed as $V=D^2/a^3$ where $D$ is the dipole moment and $a$ is the lattice spacing, such that $r_{ij}$ is the distance between sites $i$ and $j$ in units of $a$.

The molecules are confined to a 2D plane by a strong transverse trapping field, with harmonic frequency $\omega_\bot$, which prevents the molecules from collapsing from the attractive interaction between aligned dipoles.
By appropriately tuning $\omega_\bot$~\cite{buchler2007strongly,micheli2007cold}, the minimum available interparticle distance suppresses the tunnelling to already occupied sites~\cite{buchler2007three}.
Assuming that the initial configuration of the lattice has no double occupations, the molecules effectively behave as if they were hard-core bosons and therefore collisionally stable, avoiding losses in 
multiply-occupied sites due to inelastic collisions~\cite{ni2010dipolar,moses2017new} and/or so-called `sticky collisions' (whereby two molecules form a long-lived complex that is unobservable)~\cite{Mayle2013, Gregory2019}.
In the hard-core limit, the dBH model has Hamiltonian
\begin{equation}\label{eq:Ham}
 H=-J\sum_{<i,j>} b_i^\dagger b_j  + \dfrac{1}{2}\sum_{i\neq j}
 \dfrac{V}{r_{ij}^3}n_i n_j \, ,
\end{equation}
where the local Hilbert space at each site is restricted to have either zero or one boson and the on-site interaction term is dropped.
The DDI is generally understood to have long-range character; however, we note that in two dimensions, it can be considered a short-range interaction as its integral over all space converges. We will show, however, that its study using DMRG methods generally involves coupling between far-away sites.

\begin{figure}[t!] 
\centering
 \includegraphics[width=\columnwidth]{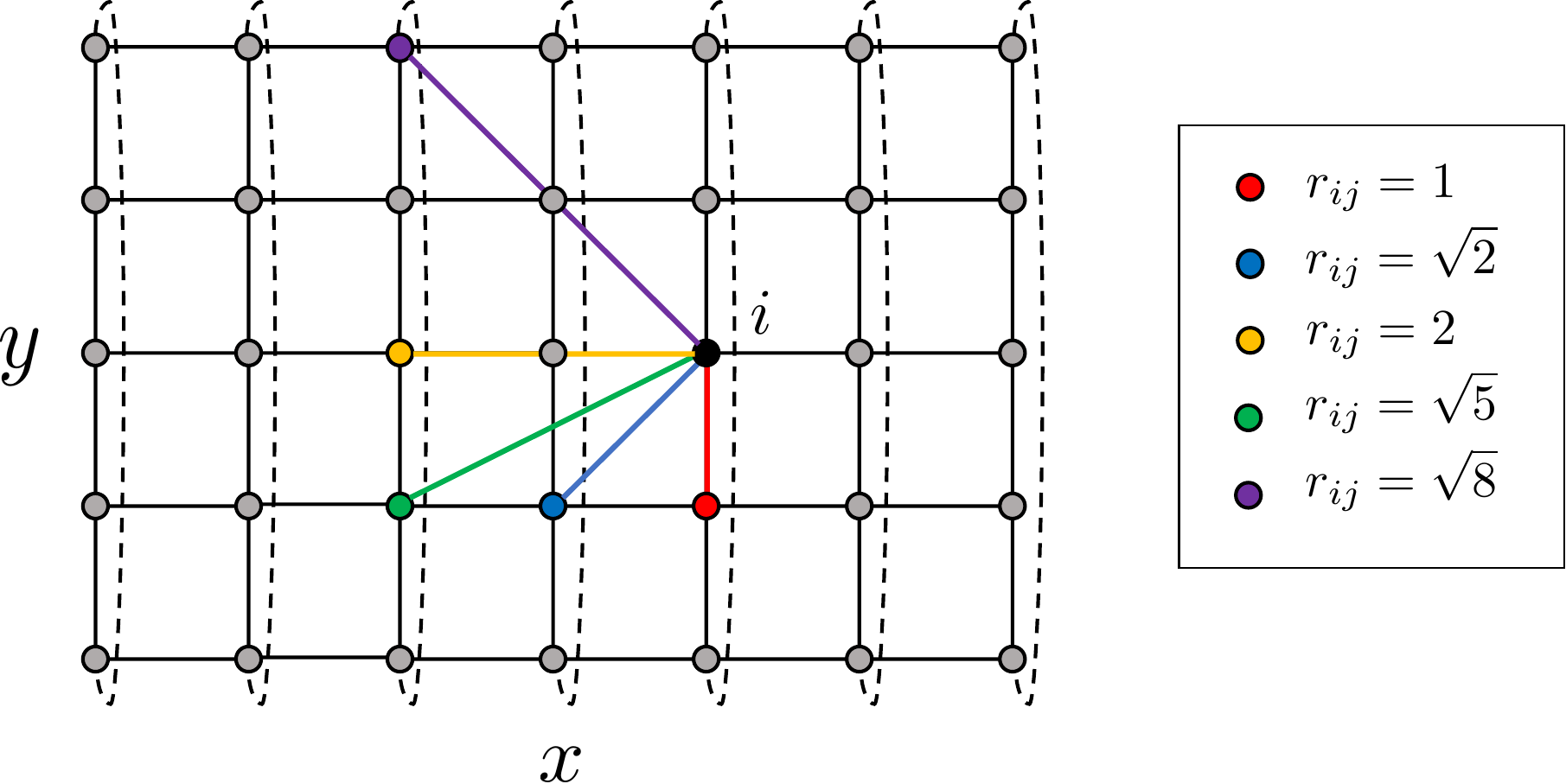}
 \caption{\label{fig:Lattice_dipoles}%
  (Color online)
  Long-range interacting terms considered in the Hamiltonian. The cutoff of the interaction is at a distance $r_{ij}=\sqrt{8}$ in units of the lattice constant; this means we include up to the $5^\mathrm{th}$ nearest neighbours. The sites are colour-coded according to their distance from the site at coordinates $i$ coloured in black.}
\end{figure}

Here, we consider the hard-core limit of the 2D dBH model in the canonical ensemble, i.e., with a fixed number of bosons, $N$, and a lattice filling $\nu=N/(L_x L_y)$, where $L_x$ is the number of sites of the lattice in the $x$-direction and $L_y$ in the $y$-direction. In our calculations, we impose periodic boundary conditions (PBC) along the $y$ direction, turning the lattice into a cylinder.
Moreover, we  truncate the range of the interaction to $r_{ij}=\sqrt{8}$ in units of the lattice spacing $a$, which is equivalent to considering interactions up to the fifth nearest neighbour [see Fig.~\ref{fig:Lattice_dipoles}].
For the case of RbCs molecules ($D\approx 1.2$ Debye) in an optical lattice with $a=532$~nm, this cut-off corresponds to neglecting interactions weaker than $\approx k_B \times 3~\text{nK}$, assuming smaller interaction effects would be washed out by the finite temperature of the sample (here $k_B$ is Boltzmann's constant).

We determine the ground state by performing DMRG calculations~\cite{DMRG} with our TNT library~\cite{TNT}. 
To this end, we map the 2D lattice to a 1D chain by sequentially going through each column of the lattice from bottom to top as in~\cite{rosson2019bosonic}.
For our model in Eq.~\eqref{eq:Ham}, this 2D-to-1D mapping turns the long-range interaction terms with range up to five nearest neighbours in 2D, to interaction terms with range up to $3L_y$ in 1D.
To efficiently treat these long-range terms, we build the matrix product operator (MPO) that describes the 2D Hamiltonian with cylindrical boundary conditions using the finite state automata technique~\cite{automata, rosson2019bosonic}. 
To guarantee number conservation, in all our calculations, we impose $U(1)$ number conservation symmetry.
For all systems considered we obtain converged results using a bond dimension $\chi$ up to 1000.

\section{Survey of order parameters
\label{sec:ord-param}}

We begin by considering the case of a small finite lattice at half filling, $\nu=1/2$, where we find only two phases, DW and SF, separated by a first-order phase transition.
We assess a set of observables to characterise solid order as a periodic density modulation (see Sec.~\ref{sec:dw}), and superfluidity as off-diagonal long-range phase coherence (Sec.~\ref{sec:sf}), as well as computing the entanglement entropy (Sec.~\ref{sec:ent-entropy}).

\subsection{Detecting periodic density modulations: static structure factor and occupation imbalance\label{sec:dw}}

\begin{figure*}[t!] 
\centering
{\includegraphics[width=1\linewidth]{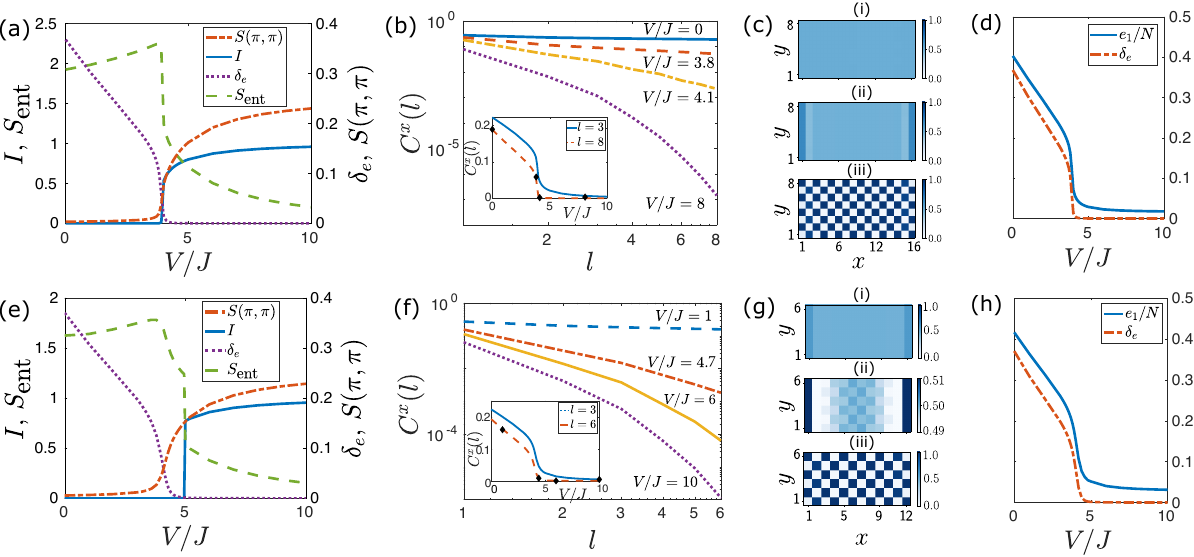}}
\caption{(Color online)
  (a) Static structure factor $S(\pi,\pi)$, occupation imbalance $I$, entanglement entropy $\Sent$, and natural occupation number difference $\delta_e$ as a function of the interaction strength $V/J$ at filling $\nu=1/2$ for $L_x\times L_y=16\times 8$.
  (b) Correlation functions along the cylinder $C^x(l)$ for the $L_x\times L_y=16\times 8$ system for the four interaction strengths indicated. The inset shows $C^x(3)$ and $C^x(8)$ as a function of the interaction strength $V/J$. The diamonds mark the four interaction strengths used in the main panel.
  (c) Local density $\tilde{n}_{x,y}$ for $V/J=0$ (i), $V/J=3.8$ (ii), $V/J=8$ (iii). Notice the different colorbar scale for panel.
  (d) Condensate fraction $e_1/N$ and $\delta_e$ as a function of $V/J$ in the $L_x\times L_y=16\times 8$ system.
  (e-h) The same quantities as in (a-d) are displayed, but for system size $L_x\times L_y=12\times 6$.
  In (g), $V/J=1$ (i), $V/J=4.7$ (ii), $V/J=10$ (iii).
  \label{fig:8x16}}
\end{figure*}

Solid order is characterized by a modulation of the density-density correlations.
In solid state physics, the structure of a crystal can be determined by X-ray or neutron scattering. The radiation impinging on a periodic density distribution will be scattered, with radiation intensity peaks located at the maxima of the static structure factor,
which is calculated as the Fourier transform of the density-density correlations as~\cite{lewenstein2012ultracold}
\begin{align}
 S(k_x,& k_y)= \nonumber \\
 = & \frac{1}{L_x^2 L_y^2}
 \sum_{x,y,x',y'}
 e^{i[ k_x(x-x')+ k_y(y-y')]} \langle n_{x,y}n_{x',y'} \rangle \,.
 \label{eq:Sk}
\end{align}
where, for the sake of clarity, we switch to the notation where the integer indices $(x,y)$ represent the site coordinates on the lattice, and $\langle \ldots \rangle$ represents expectation values with respect to the ground state.
Peaks of $S(\bm{k})$ at momenta $\bm{k}=(k_x,k_y)\neq(0,0)$ indicate the presence of solid order and a periodic modulation of density defined as $\tilde{n}_{x,y}=\langle b^\dagger_{x,y}b_{x,y}\rangle\,$ in a realization of the state.
Because of this, the structure factor has been repeatedly used to identify crystalline phases in numerical simulations of optical lattice setups~\cite{dutta2015non, ohgoe2012quantum}.
For instance, for a checkerboard solid on the square lattice at half-filling $\nu=1/2$, with the bosons localized on one of the sublattices of the square lattice [see Fig.~\ref{fig:8x16}(c)], the structure factor has non-zero peaks at $(k_x, k_y)=(\pi,\pi)$. These peaks acquire a value $S(\pi,\pi)=1/4$ in the thermodynamic limit.

The static structure factor $S(\bm{k})$ has been accessed experimentally in cold-atom setups through measurements of compressibility or density fluctuations~\cite{sanner2010suppression, drewes2016thermodynamics}.
In systems where particles are confined to move on
a discrete lattice, as our dBH model, the lattice potential provides a privileged frame from where to determine the density modulation.
Based on this, we consider here an alternative observable to detect solid order that is readily accessible in quantum gas microscope setups, the occupation imbalance~\cite{landig2016quantum,schreiber2015observation,Tai2017}.
For the checkerboard solid, the occupation imbalance is defined as
\begin{equation}
  I=\left|\frac{\sum_{x,y} (-1)^{(x+y)}\tilde{n}_{x,y}}{\sum_{x,y} \tilde{n}_{x,y}}\right|\,.
  \label{eq:dens-imbalance}
\end{equation}
It quantifies the occupation imbalance of the two checkerboard sub-lattices that make up the system: it is zero for a uniform density and 1 for a checkerboard solid,
and hence it condenses information on the whole density profile into one number. (It is straightforward to define analogous occupation imbalance functions for other crystalline arrangement, such as the stripe or star solids~\cite{Capogrosso2010}, see~\cite{RossonThesis}.)
In our numerical calculations, the two degenerate ground states of the system in the DW phase, occupying either of the sub-lattices respectively, are selectively obtained by varying the wavefunction used to initialise the DMRG calculations.

We show in Fig.~\ref{fig:8x16}(a,e) the dependence of $S(\pi,\pi)$ and $I$ on the interaction strength for $\nu=1/2$ and two system sizes, $L_x\times L_y=16\times 8$ [Fig.~\ref{fig:8x16}(a)] and $L_x\times L_y=12\times 6$ [Fig.~\ref{fig:8x16}(e)].
We find that these observables feature a similar dependence on the interaction strength. Both of them are zero or small at low $V/J$ and converge to a finite value at large $V/J$. The transition in the occupation imbalance is very sharp, with a discontinuity around the critical interaction $(V/J)_c\approx4-5$, depending on the system size.
The transition in $S(\bm{k})$ is smoother for the smaller system, but for $16\times 8$ is also very sharp, with the location of the transition in agreement with that in $I$.
The jump of  both observables at a well defined $V/J$ agrees with the expectation of the transition being of first order in the thermodynamic limit, given that two symmetries are being broken simultaneously: the $Z_2$ symmetry of the lattice is present in the checkerboard solid [$V/J > (V/J)_c$], and the $U(1)$ symmetry in the superfluid phase (see below).  
We conclude from this discussion that the occupation imbalance is able to detect the presence of a periodic density modulation as well as the structure factor, with the advantage of being readily accessible in quantum gas microscope experiments.

In addition, the occupation imbalance appears less sensitive to finite-size effects in locating the point of the transition.
Indeed, we attribute the smoother transition in $S(\pi,\pi)$ in small systems to boundary effects. We can understand this by analysing the density profiles. The insets in Fig.~\ref{fig:8x16}(g) shows the ground-state density of the $12\times6$ system for three representative values of the interaction strength. 
At $V/J=1$, in the superfluid limit, the density is uniform in the bulk.
For a large value of the interaction strength, at $V/J=10$, the density forms a checkerboard pattern.
At $V/J=4.7$, immediately below the critical $(V/J)_c\approx 4.9$ identified by the occupation imbalance, 
the central part of the lattice starts to show signs of checkerboard ordering, which are picked up by the structure factor $S(\pi,\pi)$. 
However, the difference in density of these sites is of the order of $10^{-3}$, which is negligible for the imbalance.
Fig.~\ref{fig:8x16}(c) shows the ground-state density of the larger $16\times8$ system for three representative values of the interaction strength.
At $V/J=3.8$, immediately below the critical $(V/J)_c\approx 3.9$, there is no sign of checkerboard order, indicating that the ground state is still a superfluid.
In contrast to this, for the $12\times6$ system the transition occurs at the larger value $(V/J)_c\approx 4.9$, see Fig.~\ref{fig:8x16}(g).
These values of the critical transition for finite systems should be compared with the value $(V/J)_c \approx 3.5$ obtained for the thermodynamic limit by quantum Monte Carlo methods~\cite{zhang2018equilibrium}.
These results indicate that $(V/J)_c$ is larger the smaller the system, and suggest the range of interaction strengths that experiments will need to explore to determine the actual position of the transition in a finite-size setup.

\subsection{Detecting superfluidity: momentum distribution, natural occupations, and particle correlations\label{sec:sf}}

In the previous section, we showed the dBH system is a DW solid for $V/J > (V/J)_c$. 
In this section, we discuss several observables to characterise the phase of the system for $V/J < (V/J)_c$, and find it to be superfluid.

Superfluidity is a complex phenomenon associated with a variety of properties of a system (dissipationless flow through narrow capillaries, quantized circulation, etc.).
It frequently appears entwined with Bose-Einstein condensation --- the macroscopic occupation of one single-particle state~\cite{PethickBook, PitaBook}. 
In two spatial dimensions, as we are concerned with here, the Hohenberg-Mermin-Wagner theorem rules out
condensation in the thermodynamic limit, because of thermal phase fluctuations of the order parameter~\cite{PitaBook}.
However, for finite two-dimensional systems, algebraic decay of correlations in the one-body density matrix (OBDM) is sufficient to yield a non-zero superfluid density and non-vanishing condensate fraction, in agreement with the Josephson relation between these quantities~\cite{Holzmann2007,Holzmann2007pnas}.
On account of this result, we do not compute the superfluid density or stiffness, which quantifies the energy offset caused by the introduction of a slow in-plane twist of the ground-state phase and is a measure of long-range coherence, and has been used in previous studies based on quantum Monte Carlo methods~\cite{Capogrosso2010}.

Experimentally, condensation in cold-atom lattice systems has been demonstrated by measuring the momentum distribution after release from the trap and ballistic expansion, see e.g.~\cite{Greiner2002,jimenez2010phases,landig2016quantum}, while superfluidity in two dimensions has been established, e.g., by the observation of dissipationless flow of a cold atomic gas past an obstacle moving below a critical velocity~\cite{Desbuquois2012}.
Based on these considerations, to determine the nature of the phase at $V/J < (V/J)_c$,
we calculate the one-body density matrix (OBDM), from which we access the momentum distribution and the condensation fraction.
Additionally, we compute the non-local particle correlations~\cite{Streif2016} directly from the OBDM, and evaluate their algebraic decay as a signal of superfluidity in 2D.

We compute the OBDM as
\begin{equation}
  \rho(x,y,x',y')=\langle b^\dagger_{x,y}b_{x',y'}\rangle \,,
  \label{eq:obdm}
\end{equation}
and diagonalise it to obtain its eigenvalues, $e_j$ ($j=1,\ldots,L_x\times L_y)$. These eigenvalues are also known as natural occupation numbers.
The condensate fraction of the ground state then equals $e_1/N$ and $e_1/N=\mathcal{O}(1)$ signals condensation~\cite{Penrose1956, Yang1962}. As noted above, the condensate fraction can be experimentally measured in optical lattice setups, associating $e_1$ with the population of the lowest-momentum state~\cite{Greiner2002, jimenez2010phases, landig2016quantum}; however, there is no direct experimental access to higher occupation numbers, $e_{j>1}$.

We have computed the natural occupation numbers for the ground state of the
$12\times6$ and $16\times8$ systems, and observed that the condensate fraction vanishes for $V/J>(V/J)_c$, the critical interaction strength for the appearance of DW ordering, while it reaches $\approx 0.4$ for $V \to 0$, in agreement with the findings in~\cite{Capogrosso2010}. The fact that the condensate fraction is still considerably smaller than 1 is indicative of the strong correlations in the system, due to the hard-core constraint,  which leads to a considerable depletion of the dominant single-particle state.
Numerically, we find that the difference $\delta_e=e_1/N-e_2/N$ between the two largest natural occupations has a similar behaviour to the condensate fraction, but it converges to zero more sharply on the DW side of the transition point, see Fig.~\ref{fig:8x16}(d,h).

We compare in Fig.~\ref{fig:8x16}(a,e) the dependence of $\delta_e$ on the interaction strength, with that of the structure factor and occupation imbalance, for two system sizes.
We observe in Fig.~\ref{fig:8x16}(a,e) that all three order parameters point to a first-order transition from a DW to a SF at a critical $(V/J)_c$ that only depends on the system size.
For the largest system we have simulated, $16\times8$, this is $(V/J)_c \approx 3.9$ [Fig.~\ref{fig:8x16}(a)], 
while for the $12\times6$ system $(V/J)_c\approx 4.9$ [Fig.~\ref{fig:8x16}(e)].
For both systems, these critical values agree with those obtained from the density imbalance in Sec.~\ref{sec:dw}. In general, we find that all observables point to the same critical value of $V/J$ for a each system size, with the sole exception of the structure factor, which is more sensitive to finite-size effects as explained in Sec.~\ref{sec:dw}.

We obtain the 
momentum distribution as the Fourier transform of the OBDM, $n_{(k_x,k_y)}= 1/(L_x L_y)\sum_{x,y,x',y'} \exp[i \{ k_x (x-x')+ k_y (y-y') \} ] \rho(x,y,x',y')$. 
This quantity displays a drastic change in behaviour in the following two regimes (not shown): in the SF regime, it shows a narrow peak characteristic of the condensate, whereas it gets spread out in the checkerboard DW limit.

We analyse next the decay of non-local correlations, which provide a more in-depth insight into the nature of the state of the system~\cite{PethickBook}; for a discussion on experimental methods to measure non-local density correlations we refer the reader to~\cite{Streif2016}.
In a 2D SF state, the correlation functions decay algebraically, whereas in the checkerboard solid the decay is exponential, due to the lack of long-range coherence.
We first focus our analysis on the correlation functions along the $x$-direction, 
\begin{align}
 C^x(l)=\langle b_{x,y}^{\dagger} b_{x+l,y }\rangle.
 \label{eq:dens-corr}
\end{align}
Furthermore, to minimize finite-size effects we average $C^x(l)$ over all lattice points $(x,y)$ such that both $x$ and $x+l$ are at least $L_x/4$ sites away from the boundaries, and for distances $l \leq L_x/2$.

We show in Fig.~\ref{fig:8x16}(b,f) the dependence of $C^x(l)$ on distance for four interaction strengths, representative of the transition between the SF and DW limits for the two system sizes we considered.
Because of the finite-size of the systems, and the short distances available to compute the correlation functions, it is not possible to obtain accurate exponents for their decay. 
However, using a logarithmic scale for both axes of the plot, the difference in decay behaviour becomes apparent: power-law decay looks linear, whereas exponential decay has a downward bend.
For interaction strengths smaller than $(V/J)_c$, the correlation function displays a linear behaviour on the log-log plot, which amounts to an algebraic decay, as expected for the SF regime. In contrast to this, for $V/J>(V/J)_c$, $C^x(l)$ decays fast, in agreement with the expectation for an exponential decay, in the DW phase.

The changing behaviour of the correlation functions across the transition can be illustrated more directly by fixing the distance $l$ and analysing $C^x(l)$ as a function of the interaction strength, as shown in the inset of Fig.~\ref{fig:8x16}(b,f). The two curves represent $C^x(l)$ at distances $l=3$ and $l=L_x/2$, respectively, for varying $V/J$. 
For the larger system size, in Fig.~\ref{fig:8x16}(b), we see a sharp drop in the value of the correlation function at $V/J=3.9$ which is in close agreement with all other observables we have considered [see Fig.~\ref{fig:8x16}(a)].
For the smaller system size, in Fig.~\ref{fig:8x16}(f), as a finite-size effect, the transition is smoother.

\subsection{Detecting the phase transition with the entanglement entropy\label{sec:ent-entropy}}

There is a large amount of theoretical research relating changes in the entanglement properties of ground states with the nature of correlations in the phases they support~\cite{amico2008entanglement,walsh2019local,frerot2016entanglement}. 
In particular, for systems with local interactions, the entanglement entropy (EE) upon bipartition of a system is known to satisfy scaling laws with respect to the size of the partition boundary (`area laws')~\cite{MPS,Eisert2010}, with coefficients containing information on the nature of the ground state, e.g., its topological character.
This powerful tool has been recently exploited to identify topological phases and to characterise the transition to many-body localisation in quantum-gas experiments~\cite{islam2015measuring,schreiber2015observation}.
The entanglement entropy of four-site and six-site one-dimensional (1D) Bose-Hubbard systems in an optical lattice were measured through the quantum interference of two copies of a state~\cite{islam2015measuring, Kaufman2016}. 
Ref.~\cite{schreiber2015observation} assessed the growth of entanglement entropy in larger 1D systems by measuring local density fluctuations and comparing with DMRG simulations.
Here, we calculate the EE to determine whether it can be used as an additional quantity to discriminate between the SF and DW phases.

The EE for a ground-state wavefunction is defined as the von Neumann entropy of its reduced density matrix $\rho_A$. The reduced density matrix $\rho_A$ is obtained by dividing the system into two subsets: $A$ and its complement $\overline{A}$. If the density matrix of the entire system is $\rho$, the reduced density matrix is calculated by tracing over the degrees of freedom of $\overline{A}$ to get $\rho_A  = \mathrm{tr}_{\overline{A}}\rho$.
For our system, we obtain a bipartition by cutting the cylinder along its circumference in two equal parts, each of size $L_x/2\times L_y$, with reduced density matrices $\rho_A$ and $\rho_{\overline{A}}$.
The EE of the bipartite system reads
\begin{align}
 \Sent = -\text{tr}[\rho_A\ln(\rho_A)]
       = -\sum_i |\lambda_i|^2 \ln(|\lambda_i|^2)\,.
 \label{eq:Sent}
\end{align}
The second part of the equation represents how the entanglement entropy is explicitly calculated from the Schmidt decomposition of the bipartite system where the $\lambda_i$ are the Schmidt coefficients. In our calculations, we derive the $\lambda_i$ from the matrix product form of the ground-state wavefunction, which makes them readily accessible.

We show the EE as a function of the interaction strength in Fig.~\ref{fig:8x16}(a,e) for the $16\times 8$ and $12\times 6$ systems respectively.
In both cases, we observe discontinuous behaviour of the EE at the same transition point indicated by the occupation imbalance, the condensate fraction, and the difference in natural occupations. This suggests that the EE is a useful quantity to determine the position of the phase transition in small systems.
Moreover, the fact that $\Sent$ vanishes for large values of the interaction informs us that the system in this regime has no long-range coherence. On the other hand, the finite value below $(V/J)_c$ is indicative of the long-range correlations in the ground state for weaker interactions. However, the EE cannot by itself inform us of the nature of the underlying correlations in the SF phase.

\section{Supersolidity triggered by 1-particle doping\label{sec:doped}}

An exciting prediction of Ref.~\cite{Capogrosso2010} is the existence of a supersolid phase in the dBH model around the Mott lobes, separating the DW and SF phases, in systems with a fixed chemical potential. 
In continuous systems, supersolidity appears when two continuous $U(1)$ symmetries are broken: one associated to the translational invariance of a crystalline structure, the other to a global phase of the SF state~\cite{Boninsegni2012}.
In lattice systems, it is standard to define the spontaneous breaking of translation invariance with reference to the discrete translation symmetry of the Hamiltonian~\cite{Boninsegni2012}.
Here, we analyse the detectability of SS induced by doping the checkerboard solid around half filling~\cite{Capogrosso2010}
in small lattices with a fixed number of particles.
To do this, we add an extra particle to the system with $\nu=1/2$ and observe how the increased density affects the ground state of the system, by monitoring the behaviour of the order parameters for DW and SF order discussed in Sec.~\ref{sec:ord-param}.

\begin{figure}[t!]
\centering
{\includegraphics[width=0.8\linewidth]{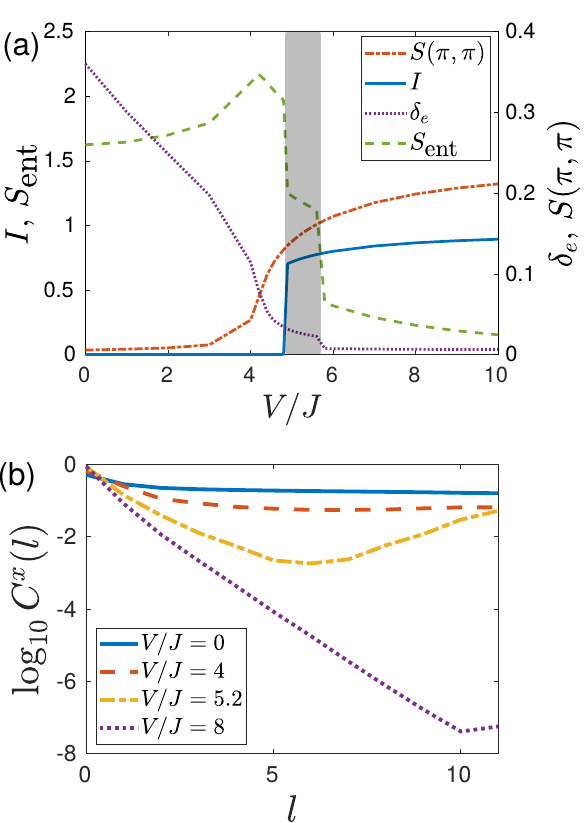}}
\caption{(Color online)
 (a) Entanglement entropy $\Sent$, occupation imbalance $I$, structure factor $S(\pi,\pi)$ and natural occupation number difference $\delta_e$ as a function of the interaction strength $V/J$ for a $L_x\times L_y=12\times 6$ cylinder at filling $\nu=37/72$.
 (b) Correlation functions along the cylinder $C^x(l)$ for four interaction strengths for the same system.  \label{fig:doped}}
\end{figure}

We show in Fig.~\ref{fig:doped}(a) the dependence of the order parameters on the interaction strength for the ground state of the $12\times6$ cylinder system with 37 bosons.
We observe that the static structure factor and occupation imbalance indicate the presence of DW order for $V/J \gtrsim 4.9$, while the difference in natural occupations, $\delta_e$, is non-zero for $V/J \lesssim 5.8$.
Combined, these results indicate the system to be a SF for $V/J \lesssim 4.9$, to support DW order for $V/J \gtrsim 5.8$, and to display supersolid (SS) properties in the region $4.9 \lesssim V/J \lesssim 5.8$.
We notice in particular the change in behaviour of $\delta_e$
in this region of interactions compared to the filling fraction $\nu=1/2$: it no longer abruptly approaches zero at the transition, but saturates at $\delta_e \approx 0.025\approx 1/37$, before becoming negligible for $V/J\gtrsim 5.8$. We interpret this behaviour as indicating the extra particle condensing `above' the checkerboard DW structure, and providing it with long-range coherence.

This two-step transition is precisely captured by the entanglement entropy, which features two sharp transitions at $V/J\approx 4.9$ and 5.8.
This demonstrates once more that the entanglement entropy is a useful quantity to determine the locations of phase transitions -- here the transition between SF and SS and between SS and DW.

The correlation functions, calculated from the left edge to the right,  show a peculiar behaviour in the SS phase [see Fig.~\ref{fig:doped}(b)].
They begin by decaying in a fashion close to exponential but show resurgence when reaching the opposite edge.
This suggests the presence of correlated modes at the edges of the system. To elucidate this point, we analyse in further detail the nature of the correlations in the ground state as a function of position, $x$, along the cylinder.

\begin{figure}[tb] 
 \centering
 \includegraphics[width=1\linewidth]{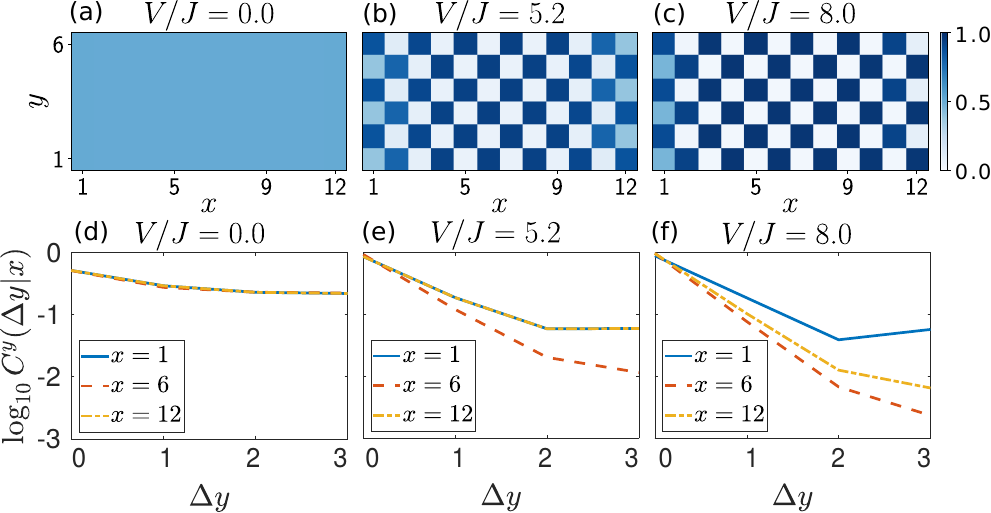}
 \caption{\label{fig:doped_correly} %
  (Color online)
  (a-c) Local density $\tilde{n}_{x,y}$ for three representative values of the interaction strength $V/J$ for a $L_x\times L_y=12\times 6$ cylinder at filling $\nu=37/72$.
  (d-f) Correlation functions along the circumference of the cylinder $C^y(\Delta y|x)$ in log scale for the same system size and interaction strengths. Each line represents a separate $x$-coordinate.}
\end{figure}

Fig.~\ref{fig:doped_correly}(a-c) shows the local density $\tilde{n}_{x,y}$ in the three regimes.
When $4.9 \lesssim V/J \lesssim 5.8$ the extra particle in the system occupies the two edge rings of the cylinder and for $V/J\gtrsim 5.8$ it occupies either one of the edge rings.
We interpret this as phase separation occurring in the system where the bulk is in the DW phase and the edge sites are in the SS phase.
The phase separation is induced by the presence of the edges that energetically favours the occupation of bosons because of the decreased interaction energy.
We expect this energetic effect to also play a significant role in experiments on finite lattices, albeit it could be reduced in the presence of harmonic trapping, an effect beyond the scope of our study.
The local density shows the presence of a density modulation throughout the system.

To confirm that ODLRO appears at the edges of the system we consider the correlation function around the cylinder in the $y$-direction at a fixed position $x$ along the cylinder,
\begin{align}
 C^y(\Delta y|x)=\langle b_{x,0}^{\dagger} b_{x,\Delta y }\rangle,
 \label{eq:dens-corr-y}
\end{align}
where, because of the PBC in $y$, $\Delta y$ goes from $0$ to $L_y/2$.
If the SS phase is limited to the edges and the bulk of the system is solid, we expect a different behaviour of this correlation function for different values of $x$.
Fig.~\ref{fig:doped_correly}(d-f) shows the behaviour of the correlation functions  $C^y(\Delta y|x)$ at the two edges of the cylinder ($x=1,12$) and in the center ($x=6$) for the ground state of the system in the SF, SS and DW regimes.
In the SF regime ($V/J < 4.9$),
$C^y(\Delta y|x)$ decays by only an order of magnitude as $y$ increases for all values of $x$, see Fig.~\ref{fig:doped_correly}(d).
For $V/J=8$ [DW regime, Fig.~\ref{fig:doped_correly}(f)], a similar behaviour of $C^y(\Delta y|x)$ is observed only at the edge of the cylinder where the extra particle is to be found ($x=1$ in these simulations). In contrast to this, the correlations along $y$ fall by over two orders of magnitude at the centre as well as the opposite edge.
In the SS regime [$V/J=5.2$, Fig.~\ref{fig:doped_correly}(e)], we observe an intermediate behaviour: the correlations decay by only an order of magnitude at the edges but fall more drastically at the centre. This indicates that in the SS regime each edge supports ODLRO and superfluidity along the $y$ direction, together with a density modulation -- i.e., each edge features SS order.
Moreover, the resurgence of $C^x(l)$ from the left edge to the right edge indicates that the two SS regions are coherently correlated, which is in agreement with 
the non-zero value of $\Sent$ and $\delta_e$ for $4.9 \lesssim V/J \lesssim 5.8$.

We explore the robustness of this SS phase by changing the doping of  the system. We consider the same system with one hole, at filling $\nu=35/72$ and with two extra bosons, at $\nu=38/72$.
In neither of these two systems do we see evidence for supersolidity in the investigated range of interaction strengths $V/J$. The set of order parameters we calculate have similar behaviour to the purely half-filled case.
The absence of supersolidity contradicts the results in Ref.~\cite{Capogrosso2010}, which predicted SS for both hole and particle doping.
We explain this discrepancy by noticing two differences in the features of our system.
First, we are considering small system sizes with definite edges where the boundary effects are significant.
Second, we are cutting the range of the interaction at five nearest neighbours.
Calculations with quantum Monte Carlo methods that similarly cut the range of the dipole-dipole interaction after a few sites~\cite{batrouni2000phase, yamamoto2012quantum} also indicate a lack of stability of the SS phase.
This suggests that SS order is rather fragile and may be washed out in lattice experiments even at nK temperatures.

Finally, we consider a system with size $L_x\times L_y=24\times6$ and $\nu=37/72$, which has the same filling as the $L_x\times L_y=12\times6$ system showing supersolidity.
We find that the local densities and correlation functions along $y$ show a similar behaviour to those in Fig.~\ref{fig:doped_correly}.
In particular, the local density shows that the two additional particles localise, for a range of values of $V/J$ around the disappearance of SF order, on opposite edges. In this regime, the correlations in the $y$-direction indicate SS order localised on the edges, similar to the $12\times 6$ system in the DW phase. 
However, the correlations in the $x$ direction show a persistent decay and no resurgence, which means there is no  coherence between the opposite edges.
This indicates that in the $24\times6$ system the two edge modes are not correlated, and ODLRO is constrained along the (periodic) $y$ direction, which is not captured in $\delta_e$ or the entanglement entropy upon cutting the cylinder along its circumference.
In agreement with this, for this system we observe no region of $V/J$ with a simultaneous non-zero structure factor $S(\pi,\pi)$ and natural occupation number difference $\delta_e$ consistent with a SS phase:
all order parameters considered point to a transition from a SF phase directly to a DW phase, analogous to Fig.~\ref{fig:8x16}(e) for lattices at half filling.

From this analysis of a larger system at the same filling, we conclude that the long-range correlation along $x$ found in the $12\times 6$ system is a finite-size effect, which may render its observation challenging in experimental setups.

\section{Detecting phases with unsupervised machine learning\label{sec:ml}}

\subsection{Detecting phases from density measurements\label{sec:ml-dens}}

In sections~\ref{sec:ord-param} and~\ref{sec:doped}, we have demonstrated the ability to identify and characterise the phases of the dBH model through the measurement and analysis of several observables, that serve as order parameters.
Here, inspired by the recently developed detection methods with single-site resolution in quantum gas microscopes,
we explore the prospects to detect all the phases above just by applying machine learning (ML) techniques to local density measurements.

Several papers have shown the power of both unsupervised and supervised ML techniques 
to recognise phases and identify phase transition points in a broad range of physical systems~\cite{wang2016discovering, hu2017discovering,liu2017self,dunjko2018machine,broecker2017machine,van2017learning,ch2017machine}. Supervised methods need to be trained on previously labelled data sets in order to draw conclusions regarding new data. Unsupervised methods, on the other hand, can provide insights into the properties of the systems without any prior knowledge of the underlying system properties.
Among the existing unsupervised methods, We have chosen to use principal component analysis (PCA) based on the spatial density data (see details below) as the outcome of the data analysis lends itself more readily to a physical interpretation than that of other clustering methods, such as $k$-means or t-SNE.
Moreover, the ability of PCA to recognise orders, symmetry breaking and even identify the transition point in quantum many-body settings has already been demonstrated~\cite{wang2016discovering, wetzel2017unsupervised}.

\begin{figure}[t] 
 \includegraphics[width=1\linewidth]{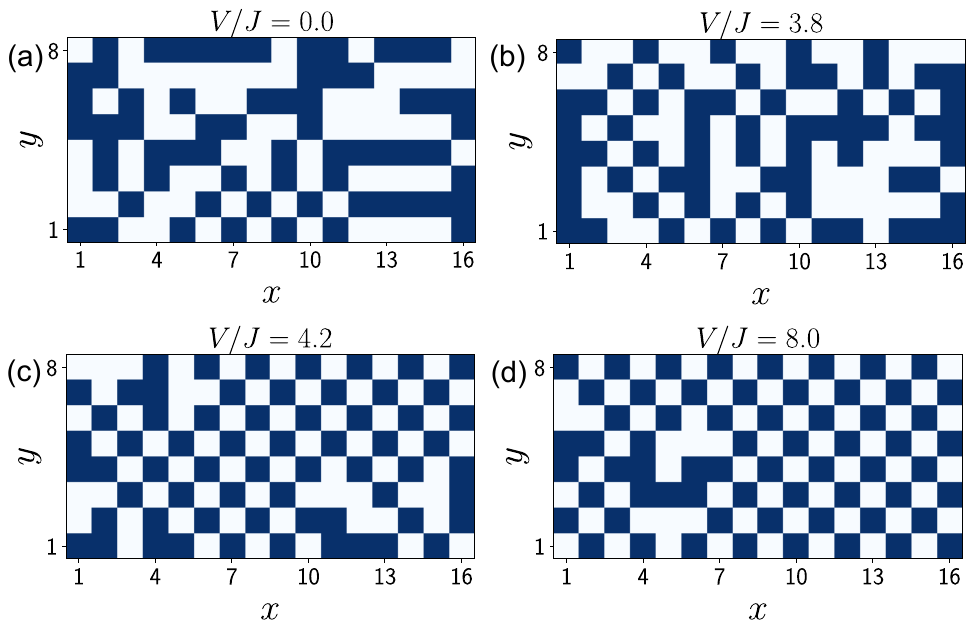}
 \caption{(Color online)
 Four instances of the sampling obtained for the $L_x\times L_y=16\times 8$ cylinder at filling $\nu=1/2$. A blue square represents a measurement of the lattice site resulted in finding a particle. A white square represents a measurement of the lattice site resulted in finding no particle. The four plots represent instances in the SF (a) and DW regime (d), and immediately before (b) and after (c) the transition.\label{fig:configs}}
\end{figure}

We apply PCA to a set of configurations of the system obtained as simulated outcomes of a single-site-resolved density measurement across the lattice.
In-situ detection is experimentally achieved by increasing the lattice depth, such that the density distribution of the gas is frozen. The parity of the occupation of the sites is then obtained through fluorescence imaging~\cite{endres2013single}.
Despite this procedure being destructive for the many-body state, the measurement captures fluctuations and correlations in the system in addition to the particle density.
For the case of hard-core bosons, the measurement of the parity is equivalent to observing the presence ($n_i=1$) or absence ($n_i=0$) of a boson in each site.

In order to numerically replicate the outcomes of this type of measurement, we consider the ground-state wavefunctions obtained from our DMRG calculations and numerically simulate a sequential projective measurement at each site as follows.
We simulate the outcome of a projective measurement resulting in the detection of a boson in the first site with probability equal to the expectation value of the number operator in the site.
We then update the ground-state wavefunction by projecting it into the subspace of the Hilbert space compatible with the measurement outcome ($n_1=0$ or $n_1=1$).
We sequentially repeat this measurement and projection procedure for all of the sites in the lattice.
This simulated measurement protocol results in an occupation pattern for the lattice which takes the correlations between particles into account as achieved in~\cite{endres2013single}.
Fig.~\ref{fig:configs} shows four sample instances of the configurations generated in this way for the $16\times8$ system with filling $\nu=1/2$ for different interactions strengths corresponding to the SF regime, the DW regime, and immediately before and after the transition. The checkerboard pattern becomes apparent after the transition, while in the SF regime site occupations are relatively random.
We generate $2000$ such particle configurations for the whole range of interaction strengths. 
For each particle configuration, we turn the set of $16\times8$ local-density values into a $128$-feature vector, and run the PCA algorithm on the resulting complete data set.

\begin{figure}[t] 
{\includegraphics[width=1\linewidth]{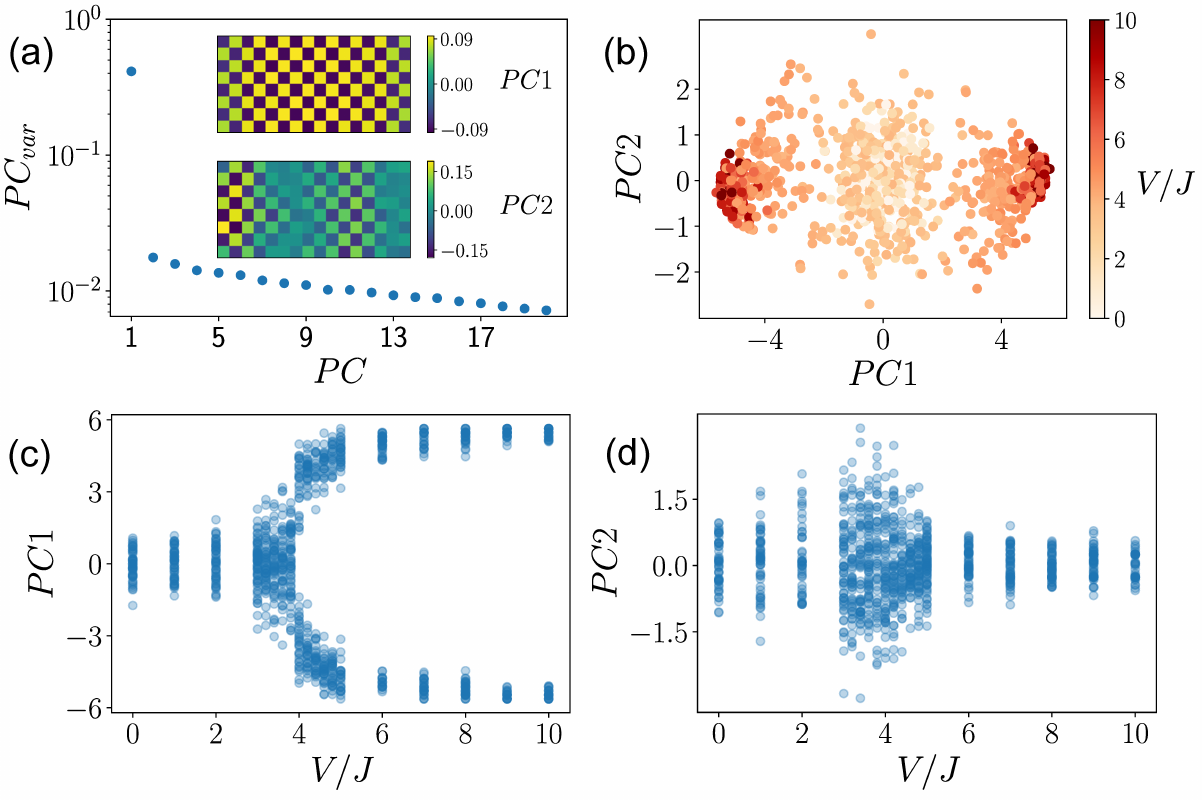}}
\caption{(Color online) Summary of the PCA analysis of the sampling for the $L_x\times L_y=16\times 8$ cylinder at filling $\nu=1/2$. 
(a) Fraction of the variance represented by each principal component. The inset shows the first two principal components. 
(b) Distribution of the instances when projected on the first and second principal components and color-coded according to the interaction strength. 
(c) First principal component as a function of the interaction strength $V/J$. 
(d) Second principal component as a function of the interaction strength $V/J$.\label{fig:PCA_tot}}
\end{figure}

We show in Fig.~\ref{fig:PCA_tot}(a) the fraction of the variance in the 
configurations data captured by each principal component.
We observe that the first principal component (PC1) accounts for $41 \%$ of the variance, 
more than one order of magnitude more relevant than the other ones.
Projecting the instances of the sampling on the first two principal components [Fig.~\ref{fig:PCA_tot}(b)]
leads to three clusters of points. The central one corresponds to the instances with a small $V/J$, and we identify it with the superfluid phase. The two other clusters represent the solid phase at large $V/J$, with each of the clusters corresponding to the particles inhabiting the two equivalent sublattices of the square lattice.
This interpretation is supported by plotting the projected coordinate on PC1 as a function of the interaction strength 
[Fig.~\ref{fig:PCA_tot}(c)], 
which displays a single branch of values $PC1\approx0$ for all configurations generated with $V/J < (V/J)_c$, which then splits into two branches, corresponding to the two $Z_2$-equivalent ground states at $V/J > (V/J)_c$.
This is further supported by the spatial representation of PC1 
[top inset to Fig.~\ref{fig:PCA_tot}(a)], which shows it essentially represents the imbalance between even and odd sites, similar the occupation imbalance $I$ in Eq.~\eqref{eq:dens-imbalance}.
The information contained in the PC1, therefore, reconstructs the occupation imbalance.
Regarding the second component (PC2), its spatial representation does not give straightforward insight into its meaning.
In agreement with this, it does not discriminate between the two phases
[cf.~Fig.~\ref{fig:PCA_tot}(d)]. 

\begin{figure}[t!] 
{\includegraphics[width=1\linewidth]{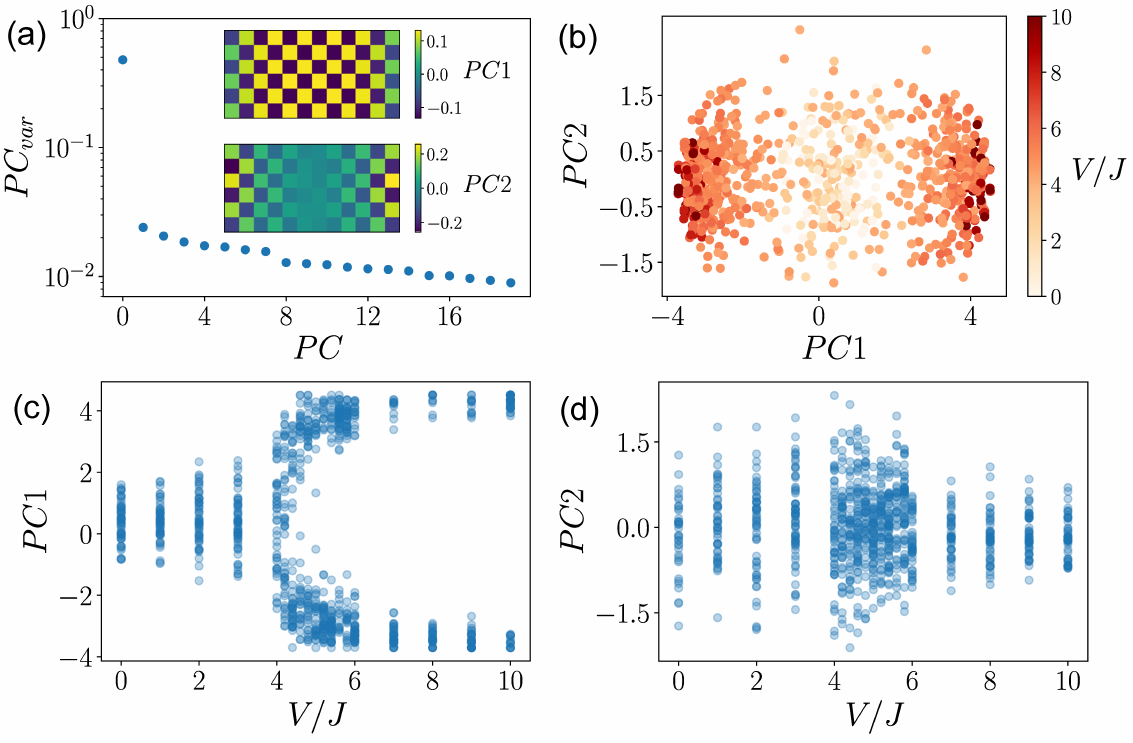}}
\caption{(Color online) Summary of the PCA analysis of the sampling for the $L_x\times L_y=12\times 6$ cylinder at filling $\nu=37/72$. 
(a) Fraction of the variance represented by each principal component. The inset shows the first two principal components. 
(b) Distribution of the instances when projected on the first and second principal components and color-coded according to the interaction strength. 
(c) First principal component as a function of the interaction strength $V/J$. 
(d) Second principal component as a function of the interaction strength $V/J$.\label{fig:PCA_doped}}
\end{figure}

We next apply the PCA approach to the doped system, to assess whether density measurements are sufficient to reveal the supersolid phase in the system with one additional particle.
We generate once again a number of simulated experimental measurements and perform a PCA in analogy with the preceding paragraph. Our results are contained in Fig.~\ref{fig:PCA_doped}.
These results are very similar to the ones obtained for the undoped system.
In particular, PC1 picks up the occupation imbalance as a discriminant between phases with and without DW order with the transition at $V/J\approx 4.9$ (cf. Sec.~\ref{sec:ord-param}).
The second principal component is dominated by boundary effects, such that an analysis of PC1 and PC2 is insufficient to reveal the existence of the SS phase.
Even though the additional particle does contribute to an increased average density and density fluctuations in the `empty' sites of the underlying checkerboard pattern, we conclude that boundary effects dominate observations in these finite, but realistically-sized, systems thus precluding the detection of the SS phase from density measurements.

\subsection{Detecting phases from entanglement properties\label{sec:ml-schmidt}}

We consider an alternative approach, based on our observations in Sec.~\ref{sec:ent-entropy}, which rely on the entanglement properties of the system.
In particular, we analyse the information contained in the Schmidt coefficients of the reduced density matrix, after a bipartition of the system [cf.~Eq.~\eqref{eq:Sent}].
Although the Schmidt coefficients are not experimentally available, they are typically obtained in numerical simulations and we apply PCA to extract from them information about the transition.
While the Schmidt coefficients do not have an immediate physical interpretation, we still resort to PCA as our unsupervised ML algorithm to analyse them for a fair comparison with the treatment of the density data in the preceding Sec.~\ref{sec:ml-dens}.
In this approach, the features making up each data sample for a given interaction strength $V/J$ are the $\chi$ largest Schmidt coefficients $\{ \lambda_n \}$ in order of decreasing magnitude.

\begin{figure}[t] 
{\includegraphics[width=1\linewidth]{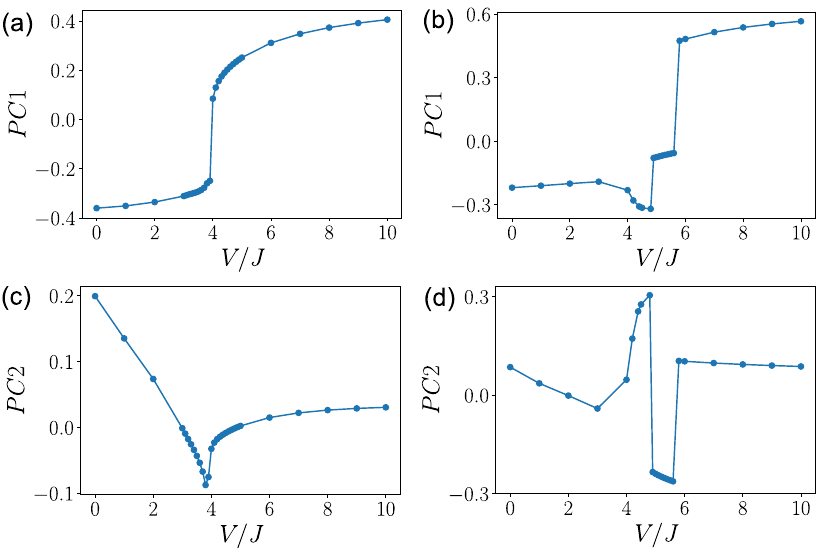}}
\caption{(Color online) (a,c) First and second principal components  as a function of the interaction strength $V/J$ for a $L_x\times L_y=16\times 8$ cylinder at filling $\nu=1/2$. (b,d) Same as (a,c) for a $L_x\times L_y=12\times 6$ cylinder at filling $\nu=37/72$. \label{fig:PCA_schmidt}}
\end{figure}

First, we consider the $L_x\times L_y=16\times 8$ system at filling $\nu=1/2$ and perform a PCA on the Schmidt coefficients, $\lambda_n$, obtained by cutting the cylinder in two equal parts. 
Fig.~\ref{fig:PCA_schmidt}(a,c) 
shows the first two principal components as a function of the interaction strength. The first PC shows a sharp change, reminiscent of the behaviour of the order parameter for a first-order transition at the same value of $V/J$ obtained from the analysis of order parameters in Fig.~\ref{fig:8x16}.
The second PC [see Fig.~\ref{fig:PCA_schmidt}(c)]
shows a cusp at the position of the transition, showing that it is sensitive to it, although we could not find an immediate interpretation of its physical content.
Repeating the same analysis for the $L_x\times L_y=12\times 6$ system doped with an extra particle, we see that the first two principal components, shown in 
Fig.~\ref{fig:PCA_schmidt}(b,d), 
manage to capture three distinct regimes. 
We notice a similarity in behaviour between PC1 and the inverse of the entanglement entropy in both systems. However, from the analysis of the $\nu=1/2$ system, we notice that PC1 grows monotonically, but the entanglement entropy does not decay monotonically. 

\section{Discussion and Summary \label{sec:conclusions}}

The use of ultracold gases in optical lattices to simulate condensed matter physical systems has allowed the exploration of previously inaccessible quantum regimes.
However, their different experimental realisation raises the question of what the best observables to characterise the phases of the system are, especially in finite-size systems available in near-future experiments.
Many numerical investigations of ultracold atoms systems make use of quantities derived from a condensed matter tradition, that however cannot always be obtained in an experimental setting.

We have performed state-of-the-art DMRG calculations to systematically evaluate and compare a set of order parameters, used in both theoretical and experimental studies,  to characterise the superfluid and solid phases of a Bose-Hubbard system with dipolar interactions.
We considered a finite system geometry with a number of particles in line with current experimental set-ups for ultracold dipolar molecules.

For a given filling fraction, we find the ground state of the system for various interaction strengths, $V/J$, and calculate a range of observables that can be used as order parameters signalling the presence of a periodic density modulation, off-diagonal long-range phase coherence, as well as the entanglement entropy. Table~\ref{Table} summarises the behaviour of all the order parameters we have considered for each phase. 

\begin{table}[tbh!] 
 \begin{tabular}{c||c|c|c|c|c|c}
 & \multicolumn{2}{c}{Solid Order}&   \multicolumn{4}{|c}{Condensation \& Superfluidity} \\ 
 \hline
 Phase & $S(\bm{k})$ & $I$ & $C^x(l)$ decay & $C^y(l|x)$ decay & $e_1,\delta_e$ & $n_k(0)$ \\
 \hline
 DW & $\neq0$ & $\neq0$ & exponential & exponential & $0$ & $\mathcal{O}(1)$ \\
 SS & $\neq0$ & $\neq0$ & -- & algebraic & $\neq0$ & $\mathcal{O}(N)$ \\
 SF & $0$     & $0$     & algebraic & algebraic & $\neq0$ & $\mathcal{O}(N)$ 
 \end{tabular}
 \caption{\label{Table}
   Phases present in the dipolar Bose-Hubbard model [Eq.~\eqref{eq:Ham}] and observables used to identify them.
   DW stands for density wave, SS for supersolid, and SF for superfluid. 
   The observables are the static structure factor at finite momentum, $S(\bm{k})$ [Eq.~\eqref{eq:Sk}], occupation imbalance, 
   $I$ [Eq.~\eqref{eq:dens-imbalance}], 
   non-local particle correlations, $C^x(l)$ [Eq.~\eqref{eq:dens-corr}] and $C^y(l|x)$ [Eq.~\eqref{eq:dens-corr-y}], 
   the natural occupations of the one-body density matrix, $e_1, \delta_e$ [Eq.~\eqref{eq:obdm}], 
   and the zero-momentum population, $n_k(0)$.
   }
\end{table}

A compelling finding of our analysis is that the set of observables that are available from the ultracold atoms toolbox is well suited to identify the phases of the system in small system sizes, and even able to indicate the presence of supersolidity.
We have shown how the occupation imbalance and the static structure factor are similarly able to capture the presence of density modulations characteristic of solid order. Moreover, we have observed that the occupation imbalance, which is readily accessible in quantum gas microscope experiments, is less sensitive to finite-size effects on small system sizes as opposed to the static structure factor, commonly used in numerical simulations.
We have observed that the condensate fraction offers a complementary order parameter to locate the SF to DW transition. Its dependence on $V/J$ for the dBH model is also consistent with the analysis of the decay of the correlation functions.

By analysing these observables, we have noted that supersolidity is very sensitive to finite-size and boundary effects and may, therefore, be challenging to detect in an experimental setting.
This sensitivity may be connected with 
recent QMC studies
of the stripe phase at $\nu=1/3$ 
which lead to opposing conclusions
on its stability in the thermodynamic limit, cf.~\cite{Bombin2017, Cinti2019}.
We rationalise these findings as due to the repulsive interaction between the particles, which favours their localisation on the edges of the system. It is possible that this energetic effect could be mitigated in systems with an overall harmonic trapping potential. We can find some evidence in this direction in recent work with a highly-magnetic atom gas in a harmonic trap (and in the absence of a lattice potential) which suggested that supersolidity was not significantly affected by finite-size effects~\cite{natale2019excitation}. It will be interesting for future research to address the competing effects of long-range interactions with trapping in the stability of a supersolid phase.

Moreover, we have shown how the entanglement entropy proves to be a versatile parameter able to pinpoint the location of the transitions, albeit not being able to offer specific insight into the coherence of the phases.

Inspired by the recent progress in using ML methods to identify phase transitions, we have employed unsupervised learning techniques on simulated experimental measurements of the occupation of single sites in the lattice.
We found that principal component analysis is able to discriminate between superfluid and solid order and naturally extracts the occupation imbalance as the relevant order parameter. However, when applied to a doped system, PCA is not able to distinguish the DW and SS phases. 
We conclude that the information available from local density measurements is not sufficient to identify the supersolid phase, and this result implies that quantities encoding long-range coherence information, such as the momentum distribution~\cite{Bottcher2019, Tanzi2019, Chomaz2019}, need to be considered to identify supersolidity.

Finally, from the PCA on the Schmidt coefficients, we conclude this approach offers valuable insight into the phase diagram of a strongly-correlated system. In particular, it may be a useful tool to locate phase transitions.
On the other hand, an analysis of physical observables across the detected phase boundaries is still required to fully unravel the nature of each phase and the particle correlations in it.

Further research should be undertaken to explore the scaling of the entanglement entropy with the system size, which can give information about the critical exponents of the phase transitions in this system.

\begin{acknowledgments}
We would like to thank J. Coulthard, Z. Hadzibabic, E. Haller, L. Santos, and H. Weimer for useful discussions.
This work has been supported by EPSRC grants No.\ 
EP/P01058X/1, 
and EP/K038311/1  
and the National Quantum Technology Hub in Networked
Quantum Information Technologies (EP/M013243/1),
and is partially funded by the European Research Council under the European Union’s Seventh Framework Programme (FP7/2007-2013)/ERC Grant Agreement No.\ 319286 Q-MAC.
We acknowledge the use of the University of Oxford Advanced Research Computing (ARC) facility in carrying out this work~\cite{ARC}.
\end{acknowledgments}

\bibliography{dipBH}

\end{document}